\documentclass{article}

\usepackage{arxiv}
\usepackage[utf8]{inputenc}
\usepackage[T1]{fontenc}
\usepackage{hyperref}
\usepackage{url}
\usepackage{booktabs}
\usepackage{amsfonts}
\usepackage{amssymb}
\usepackage{amsmath}
\usepackage{microtype}
\usepackage{graphicx}
\usepackage{subcaption}
\usepackage{enumitem}
\usepackage{tikz}
\usetikzlibrary{arrows.meta,positioning,shapes.geometric,shapes.arrows,calc}
\setlist{nosep}

\title{From Physical Devices to RTL Models: Abstraction and Validation in Hardware Engineering}
\author{Wolfgang Ecker \\
            Infineon Technologies \& \\
            TU Munich\\
            \texttt{Wolfgang.Ecker@infineon.com} \\
    \and Natalie Simson\\ 
            Infineon Technologies \& \\
            TU Munich\\
            \texttt{Natalie.Simson@infineon.com} \\    
    \and Johannes Ecker \\
            Infineon Technologies \& \\
            TU Munich\\
            \texttt{Johannes.Ecker3@infineon.com} \\       
    \and Endri Kaja\\
            Infineon Technologies \\
            \ \\
            \texttt{Endri.Kaja@infineon.com}  } 
\date{\today}

\begin{document}
\maketitle

\begin{abstract}

This paper introduces the foundational principles underlying hardware engineering models and argues that abstraction is their defining characteristic. Because abstraction necessarily omits detail and constrains what engineers can build, models are inherently incomplete in specific respects -- or, as George Box famously observed, ``All models are wrong, but some are useful'' \cite{box1979}. At the same time, abstraction is essential for simplification, which is key to managing complexity. More abstract models also tend to simulate faster because fewer details must be considered.

This paper subsequently examines a range of abstraction methods in digital design -- sometimes referred to as design disciplines -- including lumped models, value-discrete models, and time-discrete models. Together with constraints that define the validity of the abstraction and design guidelines, these abstraction methods establish design disciplines. This paper further relates these forms of abstraction to pre-clustered design elements such as transistors, gates, registers, and transfer functions. These pre-clustered elements define abstraction levels, such as the gate level, and are presented as a key enabler of increased design productivity.


\end{abstract}



\section{Introduction}
\label{sec:introduction}

Engineering would be impossible if every design decision required analysis of the complete physical system. A modern semiconductor device involves three-dimensional material structures, electromagnetic fields, nonlinear device effects, manufacturing variation, interconnect parasitics, timing behavior, logical functions, and software-visible state, among other concerns. No single representation can expose all these properties while remaining efficient enough for routine design work.

Engineers therefore work with models: selective representations that preserve the properties relevant to a particular concern while omitting others. A three-dimensional model can preserve geometry; an electrical model can preserve voltage and current behavior; a digital model can preserve logical function while ignoring continuous voltages; and an RTL model can preserve state transitions and data transfers while omitting individual gates. This selectivity is not a defect; it is the source of a model's usefulness by focusing on relevant properties.

\begin{figure} 
\centering
\begin{subfigure}[t]{0.30\linewidth}
\centering
\includegraphics[width=\linewidth]{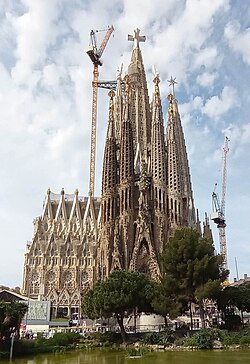}
\caption{Exterior view (web)}
\end{subfigure}
\hfill
\begin{subfigure}[t]{0.66\linewidth}
\centering
\includegraphics[width=\linewidth]{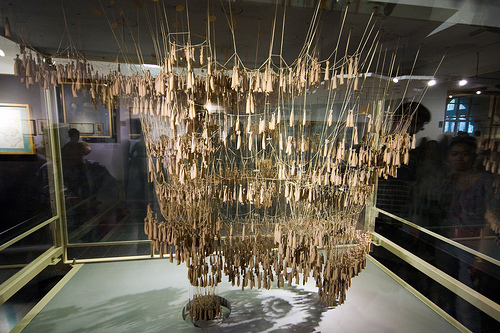}
\caption{Hanging-chain statics model (authors)}
\end{subfigure}
\hfill
\begin{subfigure}[t]{0.305\linewidth}
\centering
\includegraphics[width=\linewidth]{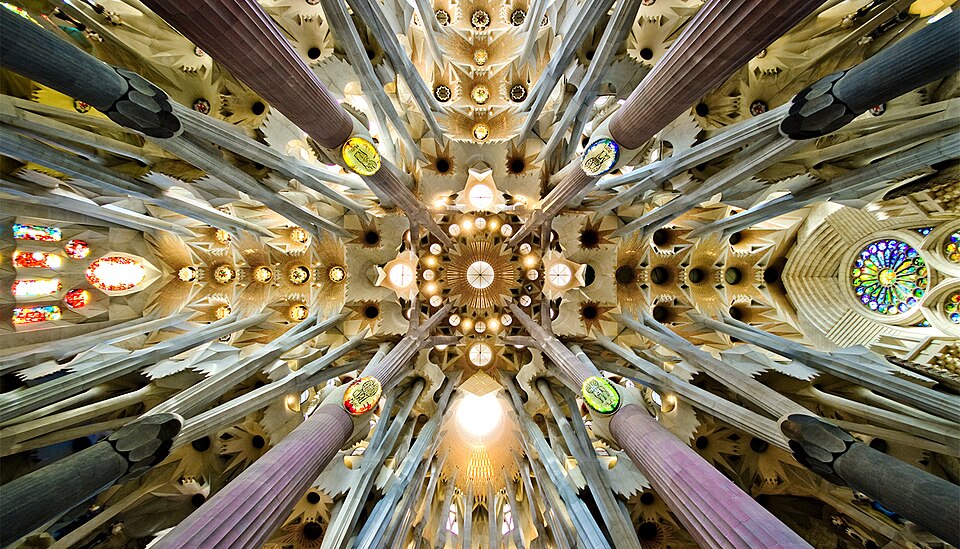}
\caption{Interior light model (web)}
\end{subfigure}
\hfill
\begin{subfigure}[t]{0.305\linewidth}
\centering
\includegraphics[width=\linewidth]{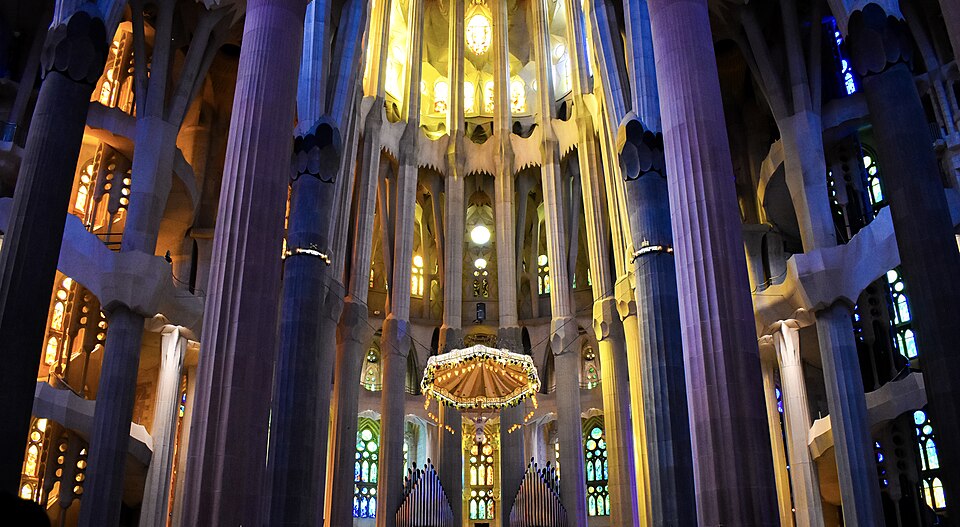}
\caption{Enlighted temple (web)}
\end{subfigure}
\hfill
\begin{subfigure}[h]{0.305\linewidth}
\centering \vspace*{-1.90cm}
\includegraphics[width=\linewidth]
{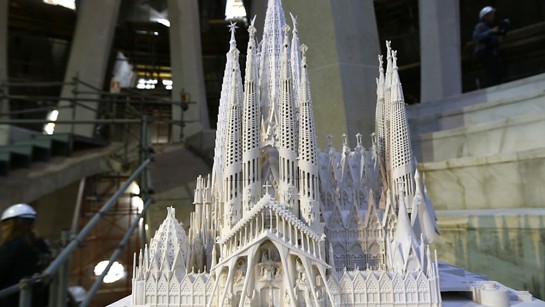}
\caption{Interior light model (web)}
\end{subfigure}
\vspace*{0.2cm}
\caption[Physical scale, exterior, and interior-light views of Sagrada Familia.]{Physical scale, exterior, and interior-light views of Sagrada Familia. \footnotesize Source: \url{https://en.wikipedia.org/wiki/Sagrada_Familia}. \vspace{-0.5cm}
}
\label{fig:sagrada-models}
\end{figure}

A building model illustrates the same principle. A physical scale model can communicate shape while saying little about structural strength. A finite-element model can predict mechanical stress while omitting color and acoustics. An acoustic model can describe sound propagation while treating much of the structure as boundary conditions. Each model represents the same building differently because it answers a different question. Figure~\ref{fig:sagrada-models} illustrates this idea using images of the Sagrada Família from Wikipedia and photographs taken by the authors.

Hardware engineering follows the same pattern and builds a stack of abstractions. Maxwell's equations provide a classical field model. Under suitable assumptions, field behavior can be reduced to lumped circuits governed by differential equations and Kirchhoff's laws. This representation can be abstracted further by replacing continuous quantities with discrete values and by describing the resulting flow of values as well as by discrete time steps and state changes.

Semiconductor structures can be represented by transistor models. Transistors can then be treated as switches and clustered into gates, while gates can be grouped into registers and combinational functions to form the register-transfer level. This hierarchy is not merely descriptive: it enables division of labor, design reuse, more efficient simulation and verification, and automated synthesis and optimization across abstraction levels.

This hierarchy also creates a recurring engineering obligation. Whenever a design moves to a higher abstraction level, engineers must show that the higher-level model preserves the properties that matter for the intended analysis or decision. This paper develops that obligation from first principles.

This paper provides a structured account of the principal abstraction levels used in digital hardware design, the transitions between them, and the principles that govern their use. It is structured as follows. Section~\ref{sec:engineering-model} establishes what a model is and what it means for a model to be appropriate for a given purpose. Sections~\ref{sec:design-disciplines-levels}, \ref{sec:physical-discipline-physical-devices}, \ref{sec:lumped-discipline-transistor-level}, \ref{sec:digital-design-discipline-gate-level}, \ref{sec:synchronous-design-discipline-rtl}, and~\ref{sec:clustering-hierarchical-design} trace the abstraction hierarchy from physical fields through lumped circuits, transistors, logic gates, and synchronous register-transfer descriptions, with attention to the design disciplines and clustering operations that enable each transition. Section~\ref{sec:hdl-moc} examines hardware description languages and models of computation as the formal substrate in which executable models are expressed. Section~\ref{sec:interface-equivalence} addresses the problem of validating the relationship between abstraction levels. Sections~\ref{sec:risks} and~\ref{sec:recommendations} identify the principal risks of model-based engineering and offer practical recommendations for managing those risks.

The paper is aimed at engineers and researchers who work with hardware models and who benefit from a clearer account of the assumptions embedded in the tools and representations they use every day.

\section{What an Engineering Model Represents}
\label{sec:engineering-model}

An engineering model is a selective representation of a real or proposed system. It captures properties judged essential for a defined set of concerns. The same system can consequently have many valid models. Three questions determine whether a model is appropriate:

\begin{enumerate}
\def\labelenumi{\arabic{enumi}.}
\item What does the model preserve?
\item What does it omit or approximate?
\item For which analyses are those omissions acceptable?
\end{enumerate}

Consider a wire on an integrated circuit. Depending on the task, it might be modeled as a three-dimensional conductor embedded in dielectric material, a distributed electromagnetic structure, a network of resistors, capacitors, and inductors, a single delay value, an ideal logical connection, or part of a timing graph.

Each representation may be correct within its intended domain and useful for selected purposes. An ideal logical wire is useful for Boolean reasoning but unsuitable for estimating crosstalk. A distributed field model can represent crosstalk but is unnecessarily expensive for checking a truth table.

Models are deliberate simplifications. Abstraction reduces information while preserving significant properties relative to a concern. The phrase ``relative to a concern'' is key. Removing geometry is reasonable for a purely Boolean analysis, but not for verifying layout-dependent timing. A useful model therefore includes an implicit contract:

\begin{itemize}
\item
  \emph{inputs and outputs}~define how the model interacts with its
environment;
\item
  \emph{state}~identifies information retained between interactions;
\item
  \emph{semantic}~defines how outputs and state evolve;
\item
  \emph{assumptions}~describe the conditions under which the model is
meaningful, i.e. provides accurate results;
\item
  \emph{guarantees}~state which properties the model promises to
preserve.
\end{itemize}

Models become dangerous when this contract remains unspecified. A simulation result may then be numerically precise but conceptually invalid because the model excludes the phenomenon responsible for the real behavior.

A modeling language and a simulator constrain different aspects of an executable model.

The~language and model of computation~determine what can be expressed and what the description means. The~simulator~implements those semantics and determines, among other things, numerical methods, event scheduling, convergence behavior, time resolution, and execution efficiency.

A detailed model can still produce misleading results if the numerical algorithm is unsuitable. Conversely, a high-quality simulator cannot recover a physical effect omitted from the model. Practical modeling limits therefore arise from both expressiveness and execution.

Scientific and engineering models are validated against observation or against more precise models rather than proven in an absolute sense. Agreement across a broad operating region increases confidence in a model, while a counterexample may reveal a missing effect, a boundary of applicability, or an inappropriate abstraction.

Maxwell's equations provide a useful example. They are fundamental to classical electromagnetism and the basis for electrical engineering and have been confirmed in countless classical electromagnetic applications, yet their domain is not unlimited. The ATLAS experiment at CERN \cite{atlas2017} illustrates this boundary: it reported high-energy light-by-light scattering in lead-ion collisions, in which photons interact and produce another pair of photons. This process is predicted by quantum electrodynamics, but it is not contained in classical Maxwell theory.

This observation does not make ordinary electrical engineering based on Maxwell's equations invalid. Rather, it identifies a regime in which the classical model is inaccurate. For typical circuit and communication applications, the neglected quantum interaction is extraordinarily unlikely to explain a device failure. Nevertheless, Maxwell's equations remain a model---and therefore an abstraction---with a defined range of applicability.

The methodological lesson is that a result outside a model's scope does not erase the model's value and accuracy within its validated operating domain. Engineers should therefore avoid asking whether a model is simply ``true'' or ``false''. The more useful question is whether it is sufficiently accurate for a specified decision under stated conditions.

\section{Design Disciplines and Design Levels}
\label{sec:design-disciplines-levels}

This paper uses two closely related concepts:~design disciplines~and~design levels. They are elaborated in the following.


A {\bf design discipline} is a deliberately imposed constraint that makes hardware design manageable \cite{agarwal2005,mit-computation-structures}. It limits the permitted structures, behaviors, or interactions so that reasoning, composition, validation, and automation become more reliable. A typical example is updating system states only at clock events

A design discipline is therefore closely related to abstraction, but it is more than a notation or drawing convention. It defines rules that make stronger guarantees possible. When those constraints are respected, the design problem becomes easier and more efficient to analyze, automate, and validate.


A {\bf design level} identifies the characteristic building blocks used to construct a design. Typical levels include physical structures; transistor models; gates and standard cells; and registers and combinational transfer functions. Moving upward in this hierarchy generally clusters many lower-level elements into a reusable higher-level component. That component exposes a simpler interface while hiding internal implementation details. Design cost and development time are reduced because engineers can rely on pre-designed and validated building blocks instead of reconstructing them from first principles.


Design discipline and design level often evolve together, they are coupled only to some extent but their distinction matters. Transistors can be studied with continuous electrical models or with simplified digital switch models. Gates can be represented as Boolean functions, timed cells, power models, physical layouts, or transistor networks. Thus, a single design level may support several views, each associated with a different analysis. Similarly, a design discipline may span several levels. This explains why hardware flows maintain multiple representations of the same component. A NAND gate may simultaneously have a lot of views. For example a transistor schematic, a Boolean truth table, an untimed logic model, a delay model, a power model, a physical layout, and a characterization model. All of these views are complementary rather than redundant.


\section{Hardware Design Disciplines and Levels}

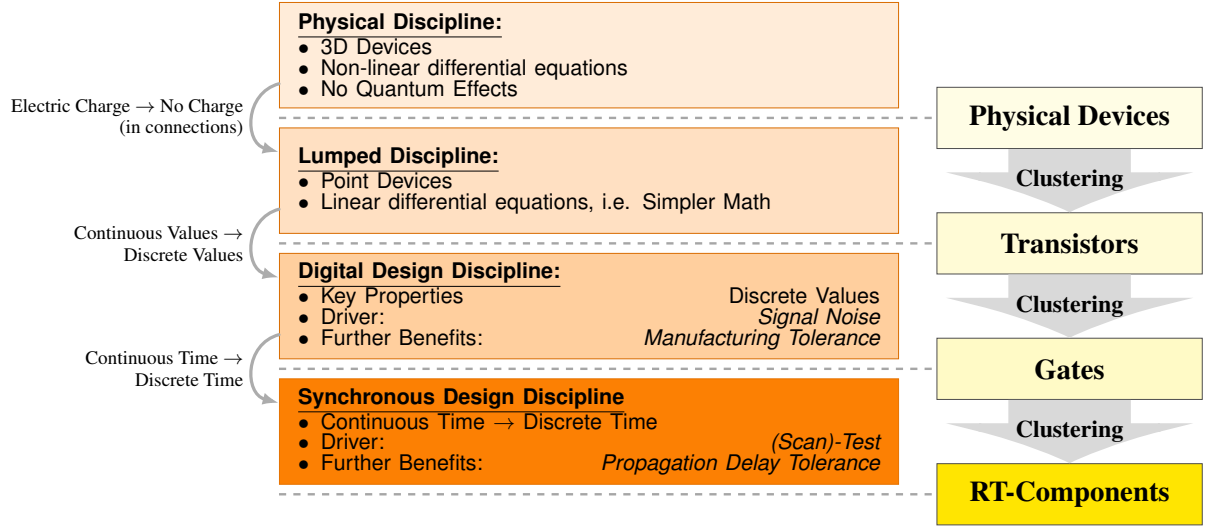
\begin{figure}
\centering
\scalebox{0.9}{ 
\begin{tikzpicture}[
	x=0.78cm,
	y=0.78cm,
	>=Latex,
	font=\sffamily,
	disciplinebox/.style={
		line width=0.4pt,
		anchor=north west,
		minimum width=9.1cm,
		minimum height=1.55cm,
		text width=8.55cm,
		align=left,
		inner xsep=0.15cm,
		inner ysep=0.04cm
	},
	rightbox/.style={
		draw=black!70,
		line width=0.4pt,
		minimum width=3.9cm,
		minimum height=0.9cm,
		anchor=west,
		align=center,
		inner xsep=0.12cm,
		inner ysep=0.00cm,
		text=black,
		font=\fontsize{12}{12}\selectfont\bfseries
	}
]
	\node[disciplinebox, draw=orange!85!black, fill=orange!15] (boxPhysical) at (4.7,9.7) {%
		{\fontsize{9}{10}\selectfont\bfseries\underline{Physical Discipline:}}\\[-0.02cm]
		{\fontsize{9}{10}\selectfont $\bullet$\ \,3D Devices}\\[-0.08cm]
		{\fontsize{9}{10}\selectfont $\bullet$\ \,Non-linear differential equations}\\[-0.08cm]
		{\fontsize{9}{10}\selectfont $\bullet$\ \,No Quantum Effects}%
	};

	\node[disciplinebox, draw=orange!85!black, fill=orange!24, below=0.28cm of boxPhysical.south west, anchor=north west] (boxLumped) {%
		{\fontsize{9}{10}\selectfont\bfseries\underline{Lumped Discipline:}}\\[-0.02cm]
		{\fontsize{9}{10}\selectfont $\bullet$\ \,Point Devices}\\[-0.08cm]
		{\fontsize{9}{10}\selectfont $\bullet$\ \,Linear differential equations, i.e. Simpler Math}%
	};

	\node[disciplinebox, draw=orange!85!black, fill=orange!38, below=0.28cm of boxLumped.south west, anchor=north west] (boxDigital) {%
		{\fontsize{9}{10}\selectfont\bfseries\underline{Digital Design Discipline:}}\\[-0.02cm]
		{\fontsize{9}{10}\selectfont $\bullet$\ \,Key Properties\hfill Discrete Values}\\[-0.08cm]
		{\fontsize{9}{10}\selectfont $\bullet$\ \,Driver:\hfill\itshape Signal Noise}\\[-0.08cm]
		{\fontsize{9}{10}\selectfont $\bullet$\ \,Further Benefits:\hfill\itshape Manufacturing Tolerance}%
	};

	\node[disciplinebox, draw=orange!90!black, fill=orange!95!brown, below=0.28cm of boxDigital.south west, anchor=north west] (boxSync) {%
		{\fontsize{9}{10}\selectfont\bfseries\underline{Synchronous Design Discipline}}\\[-0.02cm]
		{\fontsize{9}{10}\selectfont $\bullet$\ \,Continuous Time $\rightarrow$ Discrete Time}\\[-0.08cm]
		{\fontsize{9}{10}\selectfont $\bullet$\ \,Driver:\hfill\itshape (Scan)-Test}\\[-0.08cm]
		{\fontsize{9}{10}\selectfont $\bullet$\ \,Further Benefits:\hfill\itshape Propagation Delay Tolerance}%
	};

	\draw[gray!65, line width=1.35pt, line cap=round, -{Latex[length=2.4mm]}]
		([yshift=-0.42cm]boxPhysical.west)
		to[out=195,in=165,looseness=1.35]
		node[midway, left, text=black, align=right, font=\fontsize{8}{9}\selectfont] {Electric Charge $\rightarrow$ No Charge\\(in connections)}
		([yshift=0.42cm]boxLumped.west);

	\draw[gray!65, line width=1.35pt, line cap=round, -{Latex[length=2.4mm]}]
		([yshift=-0.42cm]boxLumped.west)
		to[out=195,in=165,looseness=1.35]
		node[midway, left, text=black, align=right, font=\fontsize{8}{9}\selectfont] {Continuous Values $\rightarrow$\\Discrete Values}
		([yshift=0.42cm]boxDigital.west);

	\draw[gray!65, line width=1.35pt, line cap=round, -{Latex[length=2.4mm]}]
		([yshift=-0.42cm]boxDigital.west)
		to[out=195,in=165,looseness=1.35]
		node[midway, left, text=black, align=right, font=\fontsize{8}{9}\selectfont] {Continuous Time $\rightarrow$\\Discrete Time}
		([yshift=0.42cm]boxSync.west);

	\coordinate (sepPhysical) at ($(boxPhysical.south east)!0.5!(boxLumped.north east)$);
	\coordinate (sepLumped) at ($(boxLumped.south east)!0.5!(boxDigital.north east)$);
	\coordinate (sepDigital) at ($(boxDigital.south east)!0.5!(boxSync.north east)$);
	\coordinate (sepSync) at ([yshift=-0.14cm]boxSync.south east);

	\node[rightbox, fill=yellow!12] (lvlPhysical) at ([xshift=0.55cm]sepPhysical) {Physical Devices};
	\node[rightbox, fill=yellow!20] (lvlTransistors) at ([xshift=0.55cm]sepLumped) {Transistors};
	\node[rightbox, fill=yellow!28] (lvlGates) at ([xshift=0.55cm]sepDigital) {Gates};
	\node[rightbox, fill=yellow!88!orange] (lvlRT) at ([xshift=0.55cm]sepSync) {RT-Components};

	\draw[dashed, line width=1.0pt, gray!70] ([xshift=0cm]boxPhysical.west |- sepPhysical) -- (lvlPhysical.west);
	\draw[dashed, line width=1.0pt, gray!70] ([xshift=0cm]boxLumped.west |- sepLumped) -- (lvlTransistors.west);
	\draw[dashed, line width=1.0pt, gray!70] ([xshift=0cm]boxDigital.west |- sepDigital) -- (lvlGates.west);
	\draw[dashed, line width=1.0pt, gray!70] ([xshift=0cm]boxSync.west |- sepSync) -- (lvlRT.west);

	\draw let \p1 = ($(lvlPhysical.east) - (lvlPhysical.west)$), \n1 = {veclen(\x1,\y1)}, \n2 = {0.46*\n1}, \n3 = {0.15*\n1} in
		[line width=\n2, gray!30, -{Latex[length=\n3,width=\n1]}]
		(lvlPhysical.south)
		-- node[midway, text=black, font=\fontsize{10}{10}\selectfont\bfseries] {Clustering}
		(lvlTransistors.north);

	\draw let \p1 = ($(lvlTransistors.east) - (lvlTransistors.west)$), \n1 = {veclen(\x1,\y1)}, \n2 = {0.46*\n1}, \n3 = {0.15*\n1} in
		[line width=\n2, gray!30, -{Latex[length=\n3,width=\n1]}]
		(lvlTransistors.south)
		-- node[midway, text=black, font=\fontsize{10}{10}\selectfont\bfseries] {Clustering}
		(lvlGates.north);

	\draw let \p1 = ($(lvlGates.east) - (lvlGates.west)$), \n1 = {veclen(\x1,\y1)}, \n2 = {0.46*\n1}, \n3 = {0.15*\n1} in
		[line width=\n2, gray!30, -{Latex[length=\n3,width=\n1]}]
		(lvlGates.south)
		-- node[midway, text=black, font=\fontsize{10}{10}\selectfont\bfseries] {Clustering}
		(lvlRT.north);
\end{tikzpicture}
} 
\caption{Digital Design Discipline and Design Level Stack}
\label{fig:design-disciplines-levels-overview-tikz}
\end{figure}

Figure \ref{fig:design-disciplines-levels-overview-tikz} shows the design disciplines, identified by their design and modeling abstractions, and the design levels, identified by their representative elements. While design disciplines evolve through abstraction, design levels evolve through clustering.

Each design discipline includes the abstractions of its lower-level design disciplines. Therefore, the abstraction toward discrete values in the digital design discipline also applies to the synchronous design discipline. Clustering is hierarchical as well. For example, transistors are used to build gates and are therefore also part of RT components.

The following subsections discuss the single design disciplines and design levels in detail.


\subsection{Physical Design Discipline and Physical Devices Design Level}
\label{sec:physical-discipline-physical-devices}

At the most detailed level considered here, hardware consists of three-dimensional material structures: conductors, dielectrics, doped semiconductor regions, trenches, contacts, interconnect layers, and package connections. Their behavior may involve nonlinear partial differential equations, distributed electromagnetic effects, and strong dependencies on geometry and materials.

Analyzing an entire integrated circuit directly at this level would be computationally infeasible. Such detailed field- and geometry-aware models are therefore used only where they are essential, for example in power switches, on-chip antennas, high-speed interconnects, or other structures in which distributed effects dominate. The first major abstraction is therefore the lumped-element discipline.

\subsection{Lumped Discipline and Transistor-Level Design}
\label{sec:lumped-discipline-transistor-level}

A lumped model represents a physical structure as an idealized component concentrated at a point or defined by behavior at its terminals. Familiar examples include resistors, capacitors, inductors, sources, diodes, and compact transistor models. This approximation is valid when spatial field variations and propagation effects within the component can be neglected for the analysis at hand. A common sufficient condition is that the component or circuit dimensions are much smaller than the relevant signal wavelength. Under these conditions, terminal voltages and currents adequately characterize the component's behavior. The simplification is substantial: it replaces a distributed field problem with a network problem that can be composed and analyzed using circuit laws and linear differential equations.


Lumped circuits are composed by connecting component terminals and are analyzed using Kirchhoff's laws. Kirchhoff's current law states that the algebraic sum of currents entering and leaving a node is zero:
\[\sum_{k}^{}i_{k}(t) = 0. \]
This law follows from charge conservation, expressed locally by the continuity equation derived from Maxwell's equations whereas for an ideal lumped node, charge storage at the node is neglected, so the rate of change of node charge is taken to be zero: \(\frac{\partial q}{\partial t} = 0\)
\[\oint_{}^{}{J\ ds} = \ \frac{\partial q}{\partial t}\]

Kirchhoff's voltage law states that the algebraic sum of voltage differences around any closed loop is zero:
\[\sum_{k}^{}v_{k}(t) = 0. \]
This law follows Faraday's law when time-varying magnetic flux through the loop is negligible. Under the lumped-circuit approximation, the relevant loop dimensions are small enough that induced electromagnetic effects can be ignored for the intended analysis:\(\ \frac{\partial\varnothing B}{\partial t} = 0\)
\[\oint_{}^{}{E\ dl} = \ \frac{\partial\varnothing B}{\partial t}\]

Within the lumped approximation, these laws enable systematic nodal and mesh analysis. Device equations then relate to terminal quantities. For example:
\[v(t) = R\text{ }i(t) \]
for an ideal resistor, and
\[i(t) = C\frac{dv(t)}{dt} \]
for an ideal capacitor.

A simulator such as SPICE assembles these component relations with Kirchhoff constraints into a system of equations and solves for node voltages and branch currents.


Not all lumped models are linear. Resistors and ideal sources may support linear analysis, superposition, and equivalent-circuit transformations. Diodes and transistors are inherently nonlinear over their full operating ranges. They may nevertheless be approximated through for example switch models, nonlinear compact models or small-signal models.

The choice again depends on what needs to be analyzed. A switch model may suffice for logic functionality. A nonlinear transistor model is needed for analog accuracy and timing determination of gate level cells. A small-signal model may be appropriate for gain and frequency analysis around a fixed operating point.



A {\bf physical MOS transistor} contains several semiconductor regions and field-dependent effects. When following lumped discipline, it becomes an electrical component with terminals such as gate, source, drain, and well. This abstraction performs two operations: It~clusters physical structures~into one component and it~maps physical connection regions to logical ports. The lumped discipline model no longer exposes every geometric feature. Instead, it provides equations that predict terminal currents and charges over a specified operating region.


There can be several legitimate abstractions of the same transistor which share the underlying device but preserve different properties to serve different purposes.

\subsection{Digital Design Discipline and Gate-Level Design}
\label{sec:digital-design-discipline-gate-level}

\begin{figure} 
\centering
\includegraphics[width=\linewidth]{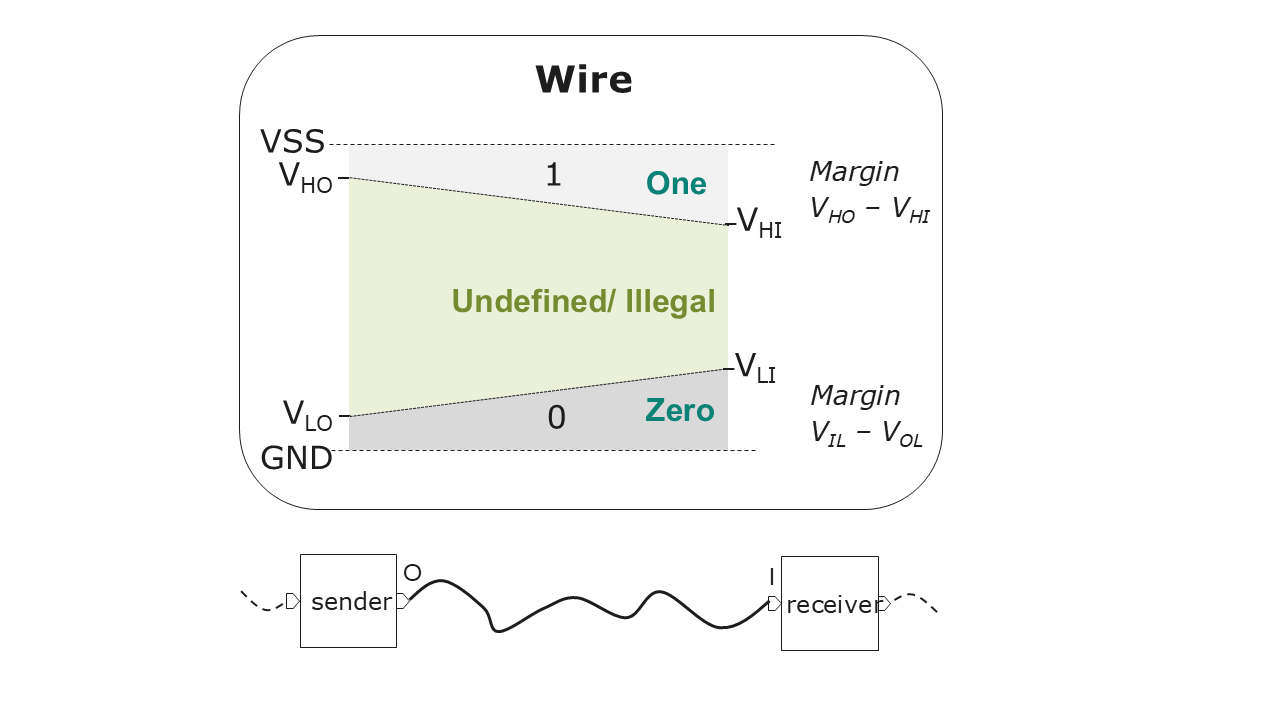}
\caption{Voltage Levels and Digital Values.
\label{fig:voltage-levels-and-digital-values}
}
\end{figure}

The digital discipline replaces continuous electrical quantities with discrete logical values. Its primary motivation is robustness. A physical wire carries a continuous voltage. As depicted in Figure \ref{fig:voltage-levels-and-digital-values} digital model classifies voltage ranges as logical zero, logical one, or invalid/undefined. A typical logic family defines:

\begin{itemize}
\item a maximum input voltage recognized as low,~\(V_{IL}\);
\item a minimum input voltage recognized as high,~\(V_{IH}\);
\item a maximum output voltage guaranteed for low,~\(V_{OL}\);
\item a minimum output voltage guaranteed for high,~\(V_{OH}\).
\end{itemize}

The differences
\[V_{IL} - V_{OL}\ \text{and }V_{OH} - V_{IH} \]
form the low and high noise margins. A valid driver produces a stronger signal guarantee than the receiver requires. This permits limited degradation without changing the interpreted logical value.

The undefined region between low and high thresholds is intentional. It marks voltages for which the digital abstraction cannot make a reliable Boolean claim.



Digital circuits are designed so that valid input levels produce valid output levels. Cascaded gates therefore restore degraded signals toward the nominal low or high regions. The implementation functions as restoring logic, with each gate contributing to this characteristic.

The digital design discipline hides many analog variations such as small voltage noise, moderate manufacturing differences, transistor parameter variation, small loading changes, as well as supply voltage and temperature variation. The result is a dramatic simplification. Instead of solving nonlinear differential equations throughout a logic network, engineers can reason with Boolean functions.


Ideal Boolean algebra has two values, 0 and 1 but these are not always enough. Executable hardware models often require additional values.

The meta-value~X~commonly represents an a voltage in the undefined regions, unknown, uninitialized, conflicting, or invalid logical value. It prevents a simulator from silently converting uncertainty into a definite zero or one.

The value~Z~represents a high-impedance or undriven state. It is important for switch-level models (see below), bidirectional buses, and nets with multiple possible drivers.

VHDL's standard logic system provides an even richer multi-value representation, including distinctions related to initialization and drive strength. SystemVerilog commonly uses four-state logic: 0, 1, X, and Z and can also support different strengths.

These meta-values remind the engineer that digital simulation remains an abstraction. An X is not a physical voltage. It represents the simulator's inability---or deliberate refusal---to select a definite Boolean value.


Under digital abstraction, a transistor can be modeled as a controlled switch:

\begin{itemize}
\item on:~the transistor creates a conductive path;
\item off:~the path is open;
\item unknown:~the controlling conditions are insufficiently defined.
\end{itemize}

\begin{figure} 
\centering
\captionsetup[subfigure]{position=top}
\begin{subfigure}[t]{0.48\linewidth}
\centering
\includegraphics[width=0.7\linewidth]{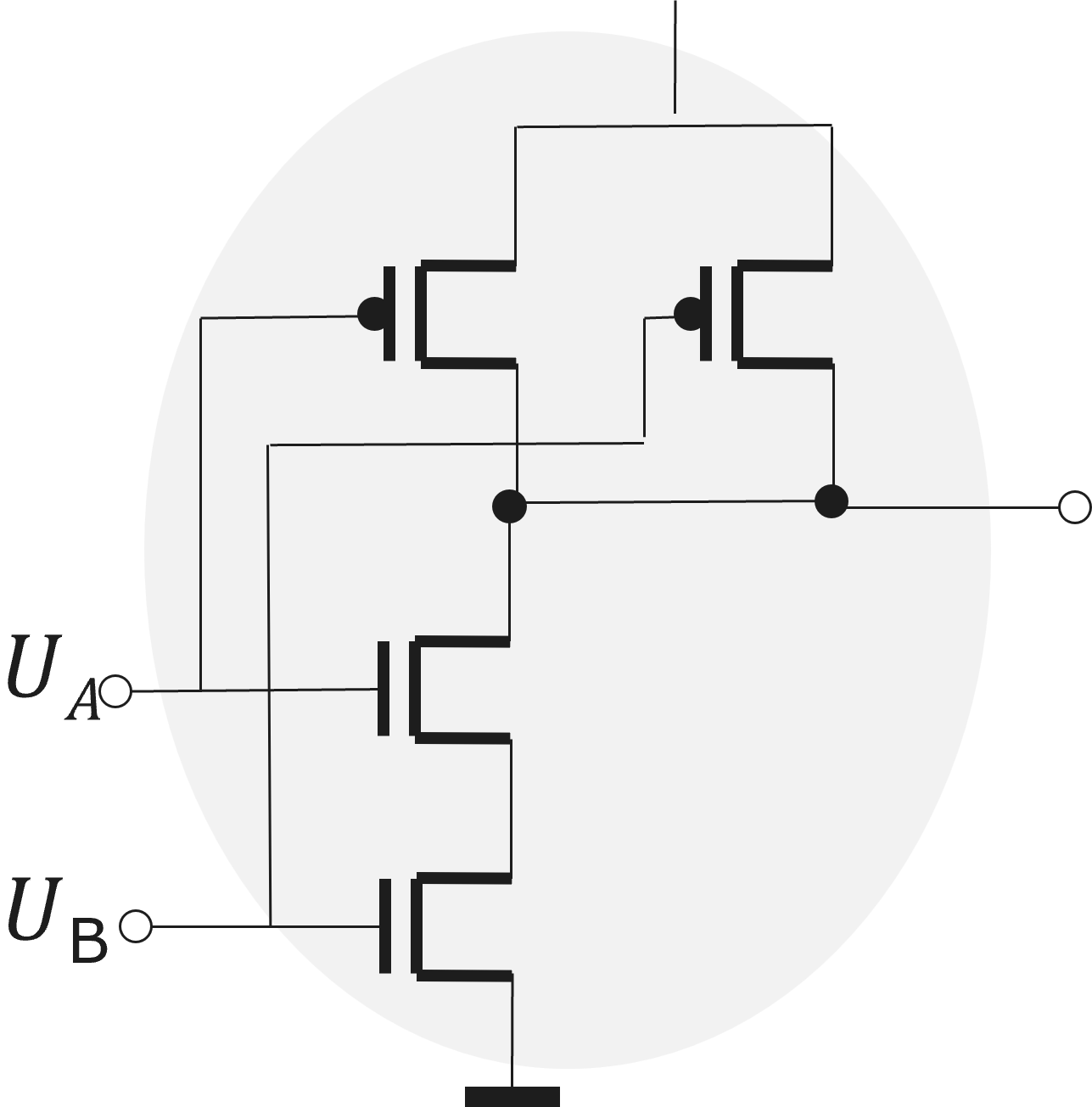}
\caption{Structural switch-level view}
\end{subfigure}
\hfill
\begin{subfigure}[t]{0.48\linewidth}
\centering
\includegraphics[width=0.8\linewidth]{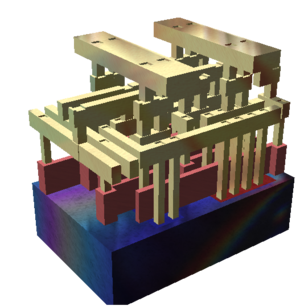}
\caption{Physical/layout-inspired realization}
\end{subfigure}
\\[0.5cm]
\hspace{0.04\linewidth}
\begin{subfigure}[t]{0.20\linewidth}
\centering
\vspace{0.4em}
\includegraphics[width=\linewidth]{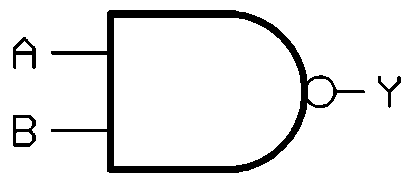}
\caption{Gate-level model}
\end{subfigure}
\hspace{0.04\linewidth}
\begin{subfigure}[t]{0.32\linewidth}
\centering
\vspace{0.0cm}
\begin{tabular}{ccc}
    \toprule
    $A$ & $B$ & $Y$ \\
    \midrule
    0 & 0 & 1 \\
    0 & 1 & 1 \\
    1 & 0 & 1 \\
    1 & 1 & 0 \\
    \bottomrule
\end{tabular}
\caption{Functional truth-table view}
\end{subfigure}
\hfill
\begin{subfigure}[t]{0.32\linewidth}
\centering
\vspace{-0.4em}
\begin{tabular}{ll}
    \toprule
    Feature & Value \\
    \midrule
    Logic function & NAND2 \\
    Rise delay ($T_r$) & 0.15 ns \\
    Fall delay ($T_f$) & 0.22 ns \\
    Update model & Event-based \\
    \bottomrule
\end{tabular}
\caption{Features of a NAND2 - Gate}
\end{subfigure}

\caption{Complementary views of a single NAND2 model across schmatic symbol, structural (transistor netlist), functional, timed, and physical representations.\footnotesize Source: Private Drawings and Tables and \url{https://en.wikipedia.org/wiki/Standard_cell}.} 
\label{fig:nand2-views} \label{fig:nand2-network}
\end{figure}

As shown in Figure~\ref{fig:nand2-network}, a transistor network can then be clustered into a logic gate. The gate hides the switch topology and exposes inputs, outputs, and a Boolean function.

Figure~\ref{fig:nand2-network} shows a transistor network forming a NAND2 gate. It realizes the following Boolean logic:
\[Y = \neg(A \land B). \]
At the gate level, this functional relationship can replace dozens of physical structures and several transistor-level equations. A transistor-level simulation in SystemVerilog can be used to validate that the transistor network in Figure~\ref{fig:nand2-network} implements the NAND function with two inputs. The meta-value Z is important in such models because it represents the effects of open transistor switches and high-impedance connections.



A gate cannot be represented by a single universal model. Figure \ref{fig:nand2-views} shows several complementary views, each answering a different engineering question. Transistor netlists and functionless, physical-realization-oriented views describe the clustering of transistors and physical elements. Truth tables represent logical functions, while timing parameters represent coarse-grained delay values. Other views may include an untimed model, which represents event ordering without physical delay; a timed model, which associates signal transitions with delays; a delay graph, which supports static timing analysis; and a power model, which estimates energy consumption or activity-dependent power.

During library characterization, more detailed transistor and interconnect models are used to derive the parameters of these higher-level views. 

~\\

\subsection{Synchronous Design Discipline and RT-Components Design Level}
\label{sec:synchronous-design-discipline-rtl}

The next major abstraction discretizes time and defines the synchronous design discipline.


In a {\bf synchronous design discipline}, state changes are coordinated by a clock. Storage elements capture new values at defined clock events, while combinational logic computes candidate next-state values between those events. The clock provides a discrete sequence $t_{0}, t_{1}, t_{2}, \ldots$. This allows designers to reason about state transitions at the RTL abstraction from cycle~\(n\)~to cycle~\(n + 1\), rather than tracking every internal analog transition and considering its detailed delay impact.

The synchronous design discipline is motivated by several practical needs: the tolerance of uncertain combinational and wire delays, predictable sequencing, scalable timing analysis, support for manufacturing testing through scan structures, and a clear separation of state from combinational computation.


Conceptually, the {\bf Register-Transfer-Level (RTL)} consists of two main elements: registers, which store state; and combinational transfer functions, which compute next values from current inputs and state. Control logic determines when and how values are selected, modified, and stored. In an RTL description, this control is expressed through the interaction of registers and transfer functions.

An N-bit register can be built from single-bit flip-flops that share a clock. Although a combinational adder is ultimately constructed from gates, its RTL representation may simply state $y = a + b$. This does not mean that the gates disappear physically. Rather, their detailed implementation is delegated to predesigned RTL components such as adders, multiplexers, and logic functions. Synthesis first translates HDL code into a technology-independent netlist composed of such generic building blocks. It then maps this netlist to a technology-specific gate-level implementation, applying optimization steps throughout the process.
 

For a simple register-to-register path, correct operation requires the clock period to accommodate the launching register delay, combinational propagation, setup time, and relevant clock uncertainty. The following {\bf timing contract} must hold:
\[T_{\text{clk}} \geq t_{\text{cq}} + t_{\text{comb}} + t_{\text{setup}} + t_{\text{uncertainty}}. \]
A hold-time constraint separately limits how quickly the new data may reach the receiving register.

The synchronous abstraction hides exact internal transition times but requires that these timing conditions are satisfied. Timing analysis therefore validates the assumptions that make cycle-level reasoning possible.


A synchronous RTL model uses discrete clock events, but the implementation remains a continuous-time physical system. Gates switch over finite intervals, wires have delays, clocks experience skew, and power supplies fluctuate.

The abstraction is useful precisely because the implementation is engineered so that these continuous effects remain within defined bounds. If the bounds fail, the abstraction may fail through setup violations, hold violations, metastability, or other timing errors.


\section{Design and Validation following Design Levels}

\subsection{Clustering and Hierarchical Design}
\label{sec:clustering-hierarchical-design}
Following design levels, groups of lower-level components are clustered into higher-level building blocks. For example, physical structures are grouped to form transistors, which are combined to create logic gates. These gates can then be used to build combinational logic components, such as adders, or sequential elements like flip-flops, which can be combined to form registers. Registers and combinational logic are then integrated to create more complex RTL components.

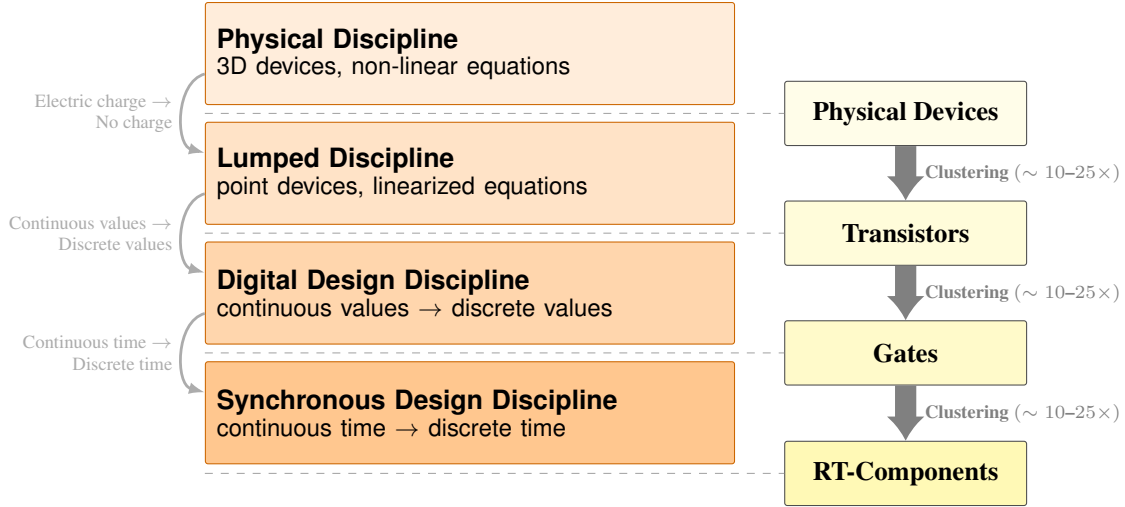
\begin{figure} 
\centering
\begin{tikzpicture}[
font=\sffamily,
>=Latex,
discbox/.style={
	draw=orange!80!black,
	rounded corners=1pt,
	align=left,
	minimum width=7.0cm,
	minimum height=1.35cm,
	text width=6.7cm,
	inner xsep=4pt,
	inner ysep=3pt
},
lvlbox/.style={
	draw=black!70,
	rounded corners=1pt,
	align=center,
	minimum width=3.2cm,
	minimum height=0.85cm,
	inner sep=3pt,
	font=\bfseries
},
node distance=0.22cm
]

\node[discbox, fill=orange!14, anchor=north west] (d1) {
	\textbf{Physical Discipline}\\[-1pt]
	\small 3D devices, non-linear equations
};
\node[discbox, fill=orange!22, below=of d1.south west, anchor=north west] (d2) {
	\textbf{Lumped Discipline}\\[-1pt]
	\small point devices, linearized equations
};
\node[discbox, fill=orange!32, below=of d2.south west, anchor=north west] (d3) {
	\textbf{Digital Design Discipline}\\[-1pt]
	\small continuous values $\rightarrow$ discrete values
};
\node[discbox, fill=orange!44, below=of d3.south west, anchor=north west] (d4) {
	\textbf{Synchronous Design Discipline}\\[-1pt]
	\small continuous time $\rightarrow$ discrete time
};

\coordinate (s1) at ($(d1.south east)!0.5!(d2.north east)$);
\coordinate (s2) at ($(d2.south east)!0.5!(d3.north east)$);
\coordinate (s3) at ($(d3.south east)!0.5!(d4.north east)$);
\coordinate (s4) at ([yshift=-0.12cm]d4.south east);

\node[lvlbox, fill=yellow!10, right=0.65cm of s1] (l1) {Physical Devices};
\node[lvlbox, fill=yellow!18, right=0.65cm of s2] (l2) {Transistors};
\node[lvlbox, fill=yellow!26, right=0.65cm of s3] (l3) {Gates};
\node[lvlbox, fill=yellow!36, right=0.65cm of s4] (l4) {RT-Components};

\draw[dashed, gray!70] (d1.west |- s1) -- (l1.west);
\draw[dashed, gray!70] (d2.west |- s2) -- (l2.west);
\draw[dashed, gray!70] (d3.west |- s3) -- (l3.west);
\draw[dashed, gray!70] (d4.west |- s4) -- (l4.west);

\draw[gray!65, line width=1.05pt, -{Latex[length=2.1mm]}]
	([yshift=-0.27cm]d1.west) to[out=195,in=165]
	node[midway, left, align=right, font=\scriptsize] {Electric charge $\rightarrow$\\No charge}
	([yshift=0.27cm]d2.west);
\draw[gray!65, line width=1.05pt, -{Latex[length=2.1mm]}]
	([yshift=-0.27cm]d2.west) to[out=195,in=165]
	node[midway, left, align=right, font=\scriptsize] {Continuous values $\rightarrow$\\Discrete values}
	([yshift=0.27cm]d3.west);
\draw[gray!65, line width=1.05pt, -{Latex[length=2.1mm]}]
	([yshift=-0.27cm]d3.west) to[out=195,in=165]
	node[midway, left, align=right, font=\scriptsize] {Continuous time $\rightarrow$\\Discrete time}
	([yshift=0.27cm]d4.west);

\draw[gray!100, line width=8pt, -{Latex[length=3.2mm,width=5.2mm]}]
	(l1.south) -- node[midway, right, font=\scriptsize\bfseries] {Clustering\ $(\sim 10$--$25\times)$} (l2.north);
\draw[gray!100, line width=8pt, -{Latex[length=3.2mm,width=5.2mm]}]
	(l2.south) -- node[midway, right, font=\scriptsize\bfseries] {Clustering\ $(\sim 10$--$25\times)$} (l3.north);
\draw[gray!100, line width=8pt, -{Latex[length=3.2mm,width=5.2mm]}]
	(l3.south) -- node[midway, right, font=\scriptsize\bfseries] {Clustering\ $(\sim 10$--$25\times)$} (l4.north);

\end{tikzpicture}
\caption{Design Disciplines and Design Levels with Clustering/Scaling Factors}
\label{fig:design-level-stack}
\label{fig:design-level-clustering-factor}
\end{figure}

The traditional stack of abstractions from physical devices to RTL is depicted in Figure~\ref{fig:design-level-stack}.


Clustering provides several benefits. It enables the reuse of proven designs and thereby accelerates development and reduces the cognitive load for designers. It also improves consistency with clearer interfaces and enables stronger automation. For example, a designer using a verified adder does not need to reconstruct its transistor network for every arithmetic operation.

Clustering also reduces flexibility. A predesigned component may be larger, slower, or more power-hungry than a special-purpose implementation. Hierarchy may introduce conservative assumptions or interface overhead. Clustering therefore trades local optimality for productivity and reliability.

As a rule of thumb, components from a higher level cluster roughly 10--25 lower-level items. This design-level clustering factor, which can be found e.g also in \cite{gerstlauer2001} is illustrated in Figure~\ref{fig:design-level-clustering-factor}.


These numbers should not be treated as physical laws. Actual ratios vary with e.g. cell complexity, technology, word width, library organization, and the exact definition of a component. A 64-bit register contains many more lower-level elements than a single flip-flop. A complex standard cell may replace several simple gates. The rule of thumb describes a multiplicative reduction in visible complexity of about one magnitude, not a fixed numerical ratio.

These clustering factors are also linked to the productivity gains achieved by using pre-clustered components, which are enabled by following the accompanying design discipline. One line of RTL therefore corresponds to approximately 10–25 gates, 100–625 transistors, and 1,000–16,625 physical items. In other words, moving up the design-level stack can deliver a productivity gain of roughly 1,000 to 16,000.
\subsection{Hardware Description Languages and Models of Computation}
\label{sec:hdl-moc}

Hardware description languages are domain-specific languages designed to represent hardware structure, behavior, concurrency, and time. (System)Verilog \cite{ieee1800} and VHDL \cite{ieee1076} are explicit hardware languages with syntax and semantics created for design and verification. SystemC \cite{ieee1666} is an embedded approach: hardware and system concepts are represented through a C++ class library and execution kernel. It is particularly useful for system-level models and hardware--software interaction. All of them are standardized.

A language’s syntax is only one part of its meaning. Its programming capabilities determine which structures and behaviors can be expressed, while its inherent model of computation (see eg. \cite{lee1998}]) defines how concurrent processes, events, values, and time interact. By combining programming capabilities with inherent models of computation, additional models of computation can be supported. For example, the events and control-flow mechanisms of hardware description languages can be used to model state machines, while libraries of places, guards, and transitions can provide a Petri-net model of computation.

A model of computation defines the semantic rules by which components execute, communicate, and update state. it determines event ordering by physical time, logical time, or another scheduling relation; multiple events potentially occurring at the same logical time; process interactions being synchronous, asynchronous, or a mixture of both; ordering and resolution of simultaneous object assignments ordered and resolved; a (non)deterministic execution as well (non)causal behavior:

Hardware simulators typically rely on discrete-event mechanisms supporting delayed events. An event queue orders future updates, while delta cycles or similar zero-time scheduling steps can resolve chains of concurrent activity without advancing simulation time.

A simulator should implement the computation model specified by the language. VHDL and (System)Verilog use different models of computation, so the chosen language must be considered together with the design discipline and associated abstraction being modeled. Different simulators for the same hardware description language may optimize scheduling differently, but conforming implementations should preserve the required observable semantics.

Simulation performance depends strongly on abstraction. A transistor-level simulation may require nonlinear numerical solution over many time steps. A gate-level simulation operates on events and logic values. An RTL simulation may execute one arithmetic operation where a gate model would process thousands of signal transitions. Higher abstraction is consequently faster, but it suppresses information. An RTL simulation cannot directly reveal all analog effects, and an untimed simulation cannot establish timing correctness.

The best model is therefore not the most detailed one. It is the least expensive model that still represents the property under investigation.

Synthesis transforms a higher-level description into a lower-level implementation. An arithmetic RTL operation may become gates; registers may become technology-specific flip-flops; logical connections may become physical wires.

The source and target need not use identical models of computation. RTL reasoning may be cycle-oriented, while the gate implementation is event-driven and physically continuous. What matters is preservation of the specified behavior under agreed assumptions. This makes synthesis a form of automated refinement rather than literal translation.

\begin{figure} 
\centering
\includegraphics[width=\linewidth]{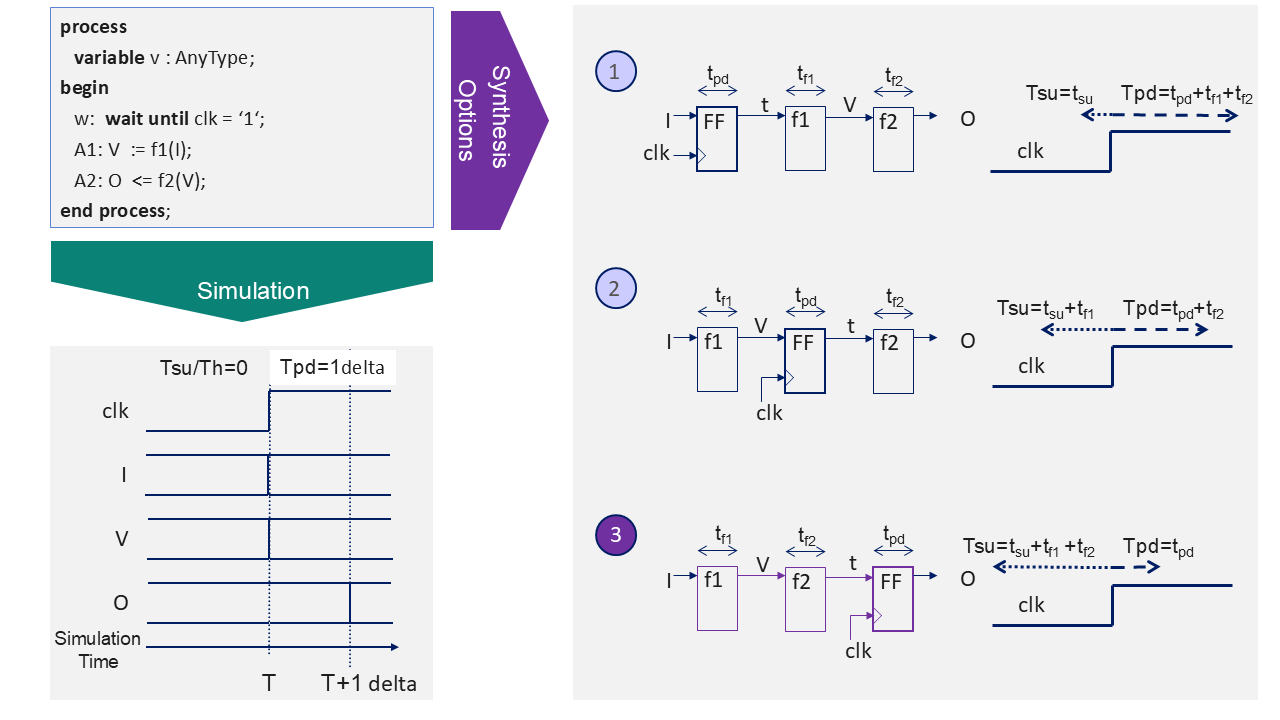}
\caption{VHDL Simulation and Synthesis Semantic of an example VHDL fragment}
\label{fig:rtl-sim-synth-semantic}
\end{figure}

Figure~\ref{fig:rtl-sim-synth-semantic} illustrates alternative representations of an example VHDL fragment. In RTL simulation, updates triggered by a clock edge are scheduled after that edge, possibly within delta cycles or a single simulation cycle. After RTL synthesis, however, the corresponding hardware is interpreted as combinational logic feeding registers before the next active clock edge. This view is more appropriate for hardware design because it makes register inputs, timing constraints, and reset behavior explicit.

\subsection{Interface Matching and Equivalence Verification}
\label{sec:interface-equivalence}

Abstraction is useful only if the relationship between levels can be trusted. Based on the relationship the correct refinement can be verified by simulation or with formal tools. Before two models can be compared, their interfaces must be aligned. At the transistor boundary, physical regions map to ports such as gate, source, drain, and body. A transistor network clustered into a gate exposes logical inputs, an output, and often implicit power connections and body connections. An adder exposes operands, a result, and possibly carry signals. A register exposes data input, data output, clock, and control signals.

Interface matching must establish the correspondence between ports, the direction, width and encoding of signals, meaning of clock and reset, power and other implicit connections, timing assumptions and legal input conditions. Comparing behavior before aligning interfaces risks proving the wrong relationship. In many cases, the preferred comparison is therefore based on externally observable behavior and avoids relying on a direct correspondence between internal states.

Formal equivalence checking, however, may require such a correspondence. It typically starts by establishing an initial-state mapping and a port mapping so that internal states and observable outputs can be compared consistently. Another case in which matching is required occurs when specifications already define so-called system states. These states are visible, either directly or indirectly, in all models and must therefore be considered during equivalence verification. If verification is limited to externally observable behavior, a reset sequence must be known to bring the system state in each model to the same value.

A higher-level component often has two important forms. A~structural implementation, composed of lower-level elements and a~functional model, describing externally visible behavior.

For a NAND gate, the structural form may be a transistor network, while the functional form is a truth table. Functional equivalence asks whether the structural implementation produces the same defined output for every legal input.

Let~\(H\)~be a higher-level model and~\(L\)~a lower-level implementation. An equivalence statement can be written conceptually as
\[\forall x \in D:obs(H,x) = obs(L,m(x)), \]
where:
\begin{itemize}
\item
\(D\)~is the set of legal inputs;
\item
\(m\)~maps the higher-level interface to the lower-level one;
\item
\(obs\)~selects the behavior relevant to comparison.
\end{itemize}

If consistent system states should be guaranteed, $x \in D$ must include a propser reset sequence.

The observation function is essential. Two models may differ in internal timing, switching activity, or intermediate values while remaining equivalent at their specified outputs and observation points.

Equivalence becomes harder when abstraction levels use different semantics. The comparison may require mapping continuous voltages to logic values or many gate events to one clock-cycle result. Further, it may need to account for latency, exclude undefined operating regions or relate different state encodings.

However, the objective is not to show that the models are identical. It is to show that the lower-level model correctly refines the properties promised by the higher-level model.

Formal equivalence checking, simulation, assertions, static timing analysis, and physical verification each address different parts of this obligation. No single technique validates every abstraction assumption.

\section{Risks of Model-Based Engineering and Practical Recommendations}
\label{sec:risks}

Models are indispensable, but their misuse can produce confident errors. Unfortunately a model may become so familiar that its assumptions disappear from view. A Boolean signal is then treated as if physical voltage no longer matters, or a cycle-accurate model is assumed to predict within-cycle timing. This confusion is particularly dangerous near abstraction boundaries.

\begin{itemize}
\item
  \textbf{Using a model outside its domain} can lead to conclusions that
are not held in practice. A lumped-circuit model, for instance, breaks down once distributed effects dominate, just as an untimed logic model cannot support a timing decision. Equally dangerous is treating X as a physical state rather than a simulation warning, the notion of a transition between 0 and 1 (or vice versa), or using typical delay values where worst-case guarantees are required.
\item
  \textbf{Ignoring uncertainty} may interpret uncertainty in a
deterministic way. However, a deterministic model may hide manufacturing variation, environmental changes, numerical uncertainty, or incomplete knowledge. Simulation precision should not be confused with confidence in model validity.
\item
  \textbf{Optimizing only the visible metric} can lead to ignoring the
invisible ones. Semiconductor design is often driven by power, performance, and area. A model optimized for one metric may omit effects that influence another. A functional RTL model, for example, may be excellent for correctness but weak for early power estimation.
\item
\textbf{Trusting refinement without validation} may lead to late bug
detection or in the worst case a bug in the final product. Automated synthesis is powerful, but every transformation relies on assumptions about legal constructs, clocking, undefined values, constraints, and tool behavior. Interface consistency and equivalence validation remain necessary even when the transformation is automated.
\end{itemize}

\label{sec:recommendations}

Before building or using a model, \textbf{state its purpose clearly}. Write down the decision the model must support. ``Model the transistor'' is vague; ``predict switching delay over the specified operating range'' is actionable.

Once the purpose is established, \textbf{identify which properties the model preserves and which it omits}. Record whether the model represents function, timing, power, geometry, or something else. Alongside this,
\textbf{document all assumptions the model relies on}, such as legal
input ranges, clock relationships, temperature limits, lumped-element conditions, initialization state, and maximum loading.

When choosing how much detail to include, always \textbf{select the lowest sufficient complexity}. Use the simplest model that preserves the target property. Additional detail increases cost and may create an illusion of accuracy without improving the decision.

Before comparing models, \textbf{validate interfaces explicitly} by checking names, widths, directions, encodings, clocking, reset behavior, implicit supplies, and observation points. Then \textbf{validate every abstraction boundary} step by step. When moving between physical devices, transistors, gates, and RTL, ask what relationship justifies each transition. Use characterization, simulation, formal checking, timing analysis, measurement, or a combination of these methods, working through each refinement step individually.

Finally, \textbf{treat unexpected results as information}. When a mismatch appears, identify the source of the unexpected result. It may indicate an implementation defect, an incorrect interface map, a tool problem, and inadequate model or operation outside the model's domain. Never assume the most detailed model is automatically correct.

\section{Conclusion}
\label{sec:conclusion}

Hardware design progresses through a sequence of deliberate abstractions. Three-dimensional physical structures are modeled as transistors; transistors are abstracted as switches and gates; gates are grouped into registers and transfer functions; and these components form RTL systems. Each step clusters lower-level elements into more capable re-usable higher-level components, improving productivity by relying on predesigned and validated building blocks. These clustered structures define the main design levels in the digital design stack: physical level, transistor level, gate level, and register-transfer level. At each level, models are abstracted thus information is removed to gain tractability. Design disciplines make this information loss safe by constraining behavior and implementation, while clustering makes it productive by enabling reuse. Hardware description languages make these models executable. Built in or library provided models of computationoffer proven concepts for abstraction. Simulators implement their computation semantics, synthesis refines high-level intent into lower-level structure, and equivalence validation checks that the relevant properties are preserved.

Design discipline and design level are distinct concepts, although they are mutually dependent to some extent. For example, a synchronous design discipline enables register-level components and supports automatic test-pattern generation as well as formal equivalence checking against gate-level components.

The central engineering principle is simple: use models deliberately and effectively, but remain aware of what they omit. Models are not substitutes for reality; they are the principal tools engineers use to understand, design, analyze, and validate it. As George Box observed,
\emph{\textbf{``Essentially, all models are wrong, but some are
useful.''}}

\section{Disclaimer}
\label{sec:references}
This paper is based on the 2026 FDL \cite{FDL2026DesignLevelAndDiscipline} and MECO \cite{MECO2026DesignLevelAndDiscipline} tutorial given by the authors of this paper. ChatGPT has been used to rework the text. Github Copilot has been applied to translate the text from word format and the pictures and drawings from powerpoint format. The content has explicitly been provided by the authors and the result reviewed.

\section{Acknowledgments}
\label{sec:acknowledgements}
This work was partly funded by the German Federal Ministry of Research, Technology and Space (BMFTR) within the project Scale4Edge under contract number 16ME0122K and project DI-FITS under contract number 16ME1168

\bibliographystyle{plain}
\bibliography{references}

@misc{FDL2026DesignLevelAndDiscipline,
  author = {Ecker, Wolfgang},
  title = {On Languages, Models, and Modelled Items},
  howpublished = {Embedded tutorial at FDL 2026 Conference and Summer School},
  year = {2026}
}

@misc{MECO2026DesignLevelAndDiscipline,
  author = {Simson, Natalie and Ecker, Johannes and Kaja, Endri and Ecker, Wolfgang},
  title = {Systematic Digital Design: Boosting Design Productivity and Component Reuse through Handshakes and Interfaces},
  howpublished = {Tutorial at MECO 2026 Conference},
  year = {2026}
}

@incollection{box1979,
  author = {Box, G. E. P.},
  title = {Robustness in the Strategy of Scientific Model Building},
  booktitle = {Robustness in Statistics},
  editor = {Launer, R. L. and Wilkinson, G. N.},
  publisher = {Academic Press},
  year = {1979}
}

@article{lee1998,
  author = {Lee, Edward A. and Sangiovanni-Vincentelli, Alberto},
  title = {A Framework for Comparing Models of Computation},
  journal = {IEEE Transactions on Computer-Aided Design of Integrated Circuits and Systems},
  volume = {17},
  number = {12},
  year = {1998}
}

@article{atlas2017,
  author = {{ATLAS Collaboration}},
  title = {Evidence for Light-by-Light Scattering in Heavy-Ion Collisions with the ATLAS Detector at the LHC},
  journal = {Nature Physics},
  year = {2017}
}

@book{agarwal2005,
  author = {Agarwal, Anant and Lang, Jeffrey H.},
  title = {Foundations of Analog and Digital Electronic Circuits},
  publisher = {Morgan Kaufmann},
  year = {2005}
}

@misc{mit-computation-structures,
  author = {{MIT}},
  title = {Computation Structures},
  howpublished = {Course materials on digital abstraction and hierarchical digital-system design}
}

@misc{ieee1076,
  author = {{IEEE}},
  title = {IEEE Std 1076-2019: VHDL Language Reference Manual},
  year = {2019}
}

@misc{ieee1800,
  author = {{IEEE}},
  title = {IEEE Std 1800-2023: SystemVerilog - Unified Hardware Design, Specification, and Verification Language},
  year = {2023}
}

@misc{ieee1666,
  author = {{IEEE}},
  title = {IEEE Std 1666-2023: Standard SystemC Language Reference Manual},
  year = {2023}
}

@book{gerstlauer2001,
  author = {Gerstlauer, A. and Dömer, R. and Peng, J. and Gajski, D.},
  title = {System Design: A Practical Guide with SpecC},
  publisher = {Kluwer Academic Publishers, Boston},
  year = {2001}
}

\end{document}